\documentclass[twoside]{ilcws08}
\usepackage[latin1]{inputenc}
\usepackage[dvips]{graphicx,epsfig,color}
\usepackage{wrapfig,rotating}
\usepackage{amssymb,amsmath,array}

\begin{document}
\title{
Background Studies for the VTX Geometry Optimisation} 
\author{Pawe{\l} {\L}u{\.z}niak
\thanks{This work was partially supported by the Ministry of Science
and Higher Education, research project no.~EUDET/217/2006 (2006-2009).}
\vspace{.3cm}\\
University of {\L}{\'o}d{\'z} - Faculty of Physics and Applied Informatics\\
 Pomorska 149/153, PL-90-236 {\L}{\'o}d{\'z} - Poland
}

\maketitle

\begin{abstract}
Influence of Vertex Detector (VTX) geometry on jet flavour tagging
performance was studied in presence of $e^+e^-$ pair background
from beamstrahlung in the International Linear Collider (ILC).
Five layer ``long-barrel'' VTX geometry with varying layer
thickness, space resolution and radius of the first layer was
tested with fast simulation tools. Influence of VTX geometry and
$e^+e^-$ pair background on measurement of SM and~MSSM Higgs boson
branching ratios was also studied.
\end{abstract}

\section{Simulation}
The $e^+e^-$ pair background from beamstrahlung was simulated with Guinea-Pig~\cite{guineapig}
for nominal accelerator parameters at 500 GeV with 14 mrad crossing angle.
The ILC detector response was modelled with Simulation \'a Grande Vitesse
2.30 (SGV).
A 5 layer ``long barrel'' VTX was considered, with the following basic configuration: $100$ $\mu$m layer thickness, $4$~$\mu$m track spatial resolution
and layer radii listed in~Table~\ref{vtxradius}.
Different sets of VTX geometries, differing from the basic configuration with value of only one parameter, were considered:
layer thickness in the range of $50-300$ $\mu$m, spatial resolution in the range $2-8$~$\mu$m and radius of the first layer in the range
$8-18$~mm. Also configuration with innermost layer removed was considered (radius of the first layer equal to 26 mm).


\begin{wraptable}{r}{0.5\columnwidth}
\centerline{\begin{tabular}{|c|c|c|}
\hline \textbf{Layer} & \textbf{Radius} & \textbf{\# of ladders}  \\
\hline 1 & 15 mm & 8 \\
\hline 2 & 26 mm & 22 \\
\hline 3 & 37 mm & 32 \\
\hline 4 & 48 mm & 40 \\
\hline 5 & 60 mm & 50 \\
\hline
\end{tabular}}
\caption{VTX detector layer specifications.}
\label{vtxradius}
\end{wraptable}
A beryllium beampipe of 0.25 mm thickness and radius 1 mm smaller than
the first layer of the VTX was assumed.
 Dedicated software was developed to track charged
particles through the VTX, taking into account multiple scattering and energy
loss. VTX was assumed to be read out 20 times per bunch train, implying that
 background hits accumulated over 131 bunch crossings and overlayed
``physics'' hits.
Tracks detected in the central tracker were refitted with hits from the VTX
(both physics and background hits), selected with the Kalman Filter (own software tools).
Jet flavour tagging was performed with ZVTOP \cite{zvtop}.
Physics processes (eg. Higgs production) were simulated with PYTHIA \cite{pythia}.

\section{Jet flavour tagging performance}
\begin{figure}[h]
\centerline{
\begin{minipage}{2mm} \vspace*{-5.5cm} a)\end{minipage}
\includegraphics[width=0.32\columnwidth]{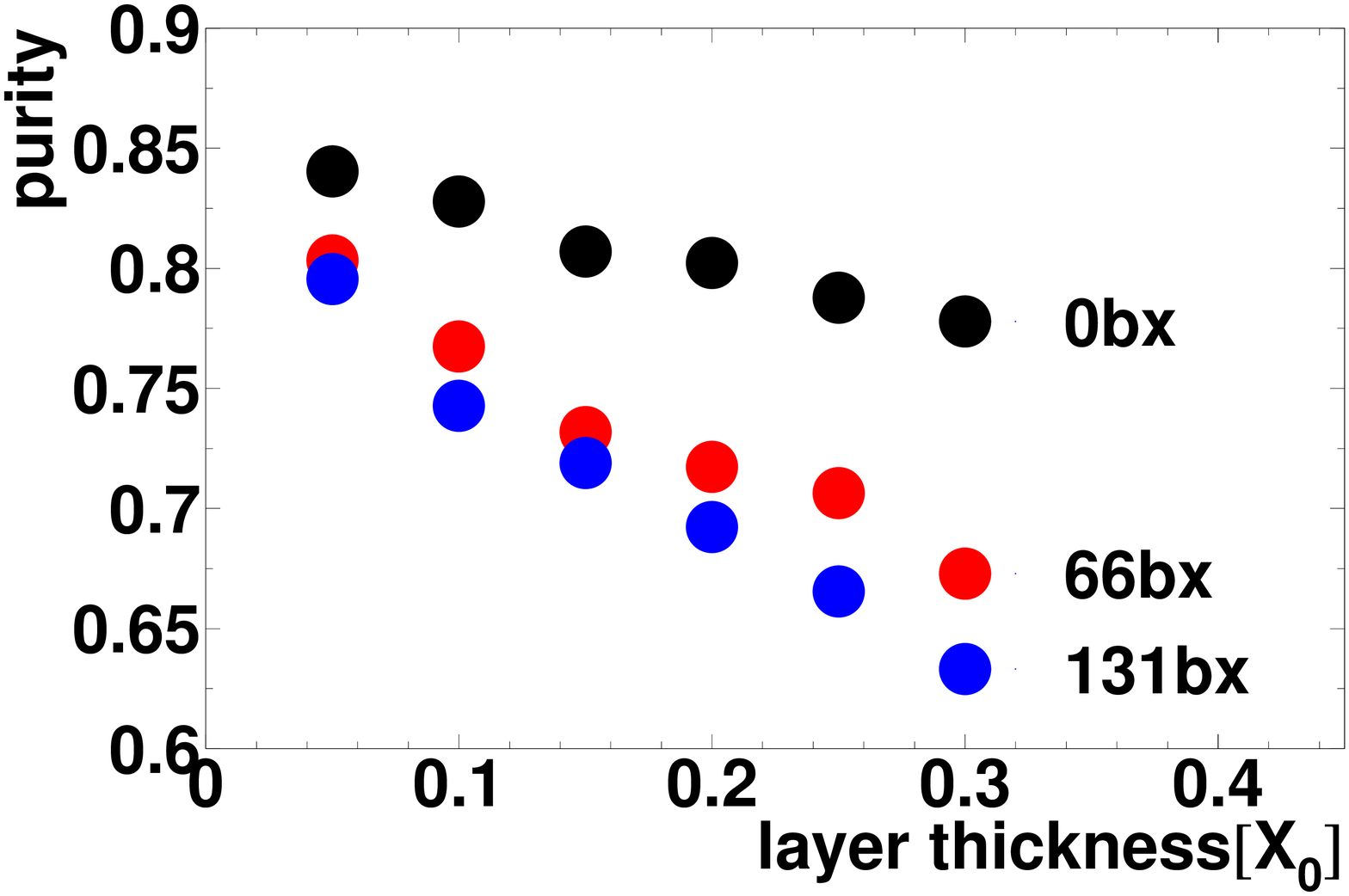}
\begin{minipage}{2mm} \vspace*{-5.5cm} b)\end{minipage}
\includegraphics[width=0.32\columnwidth]{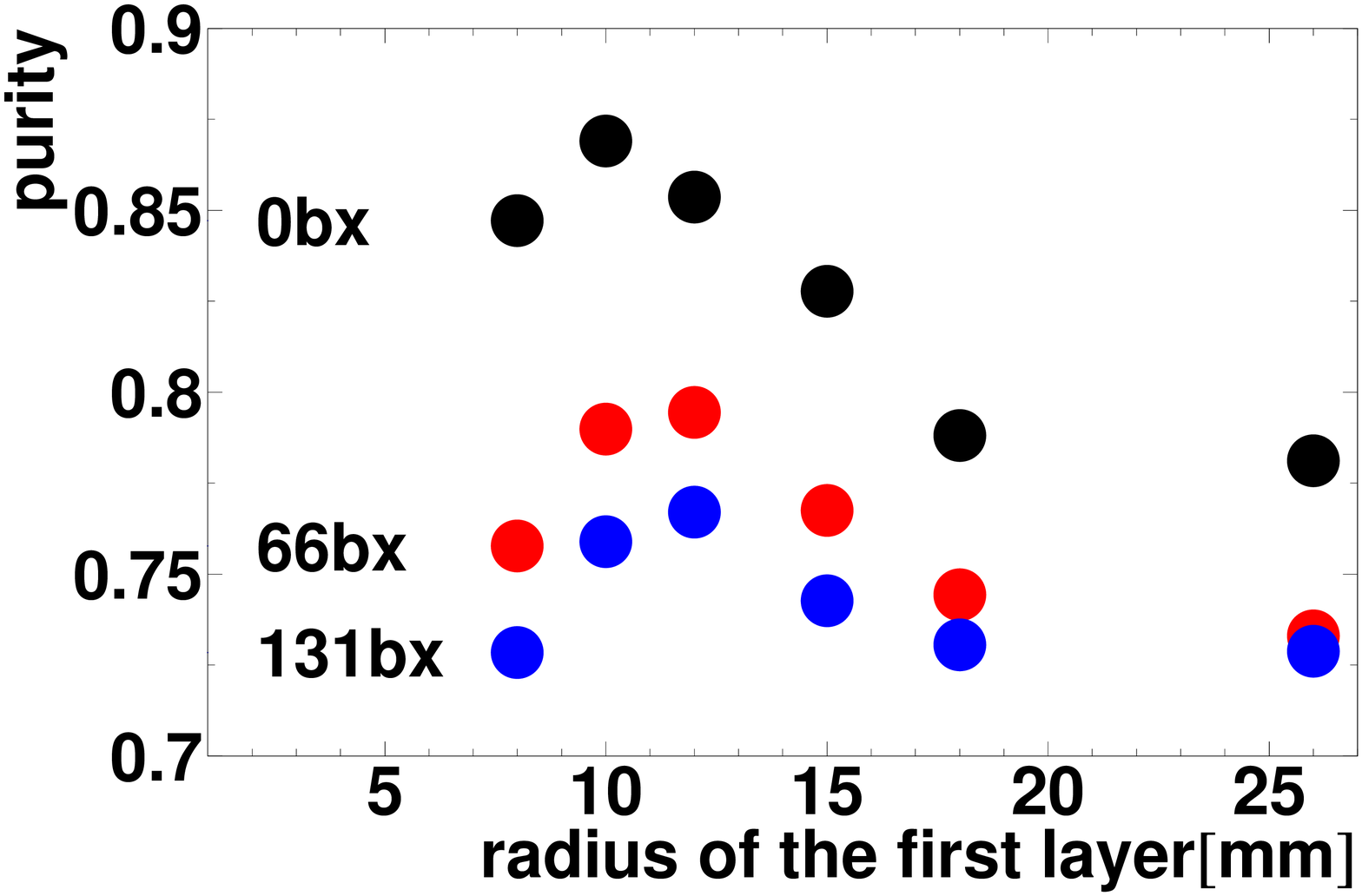}
\begin{minipage}{2mm} \vspace*{-5.5cm} c)\end{minipage}
\includegraphics[width=0.32\columnwidth]{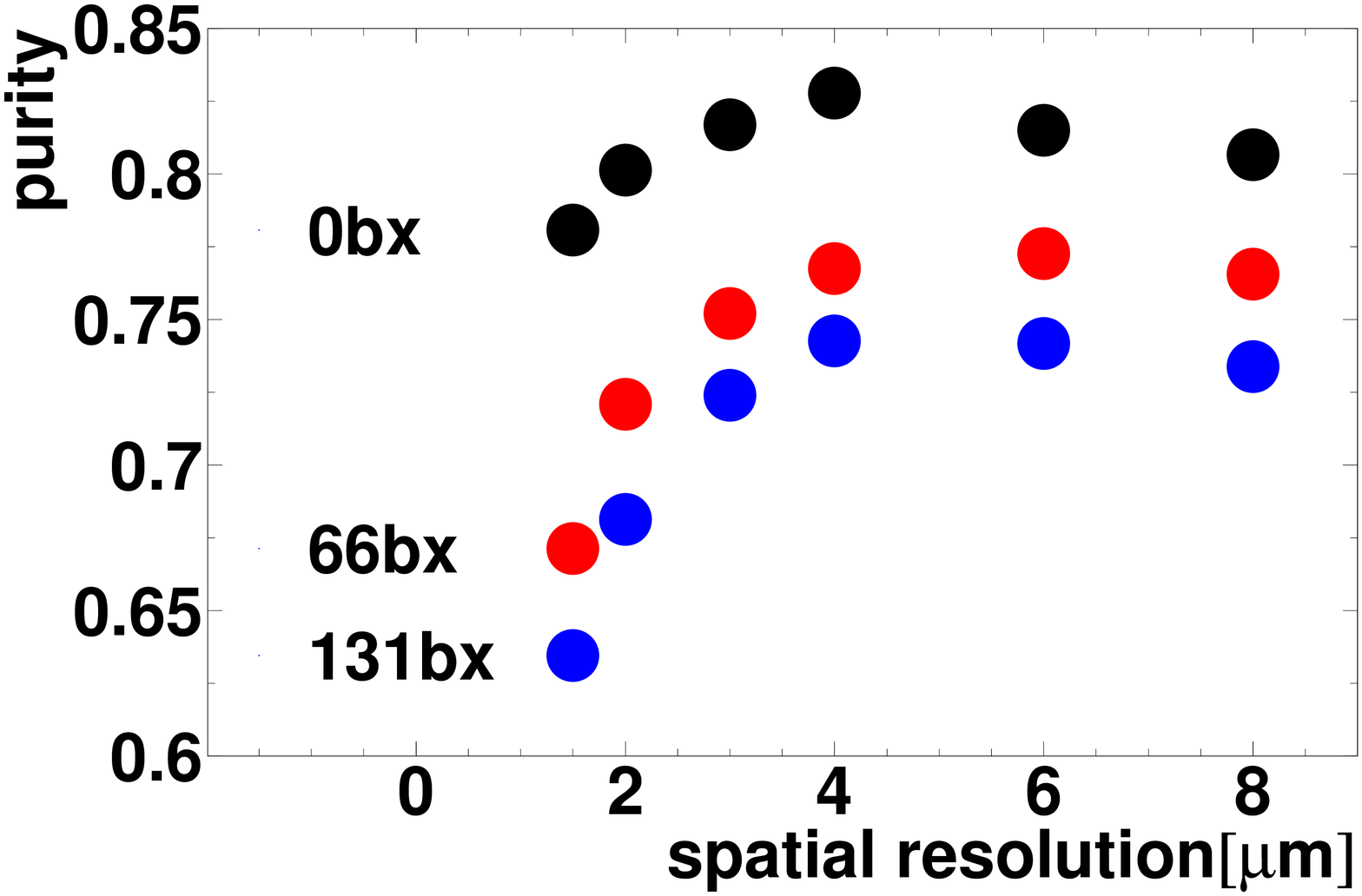}
}
\centerline{
\begin{minipage}{2mm} \vspace*{-5.5cm} d)\end{minipage}
\includegraphics[width=0.32\columnwidth]{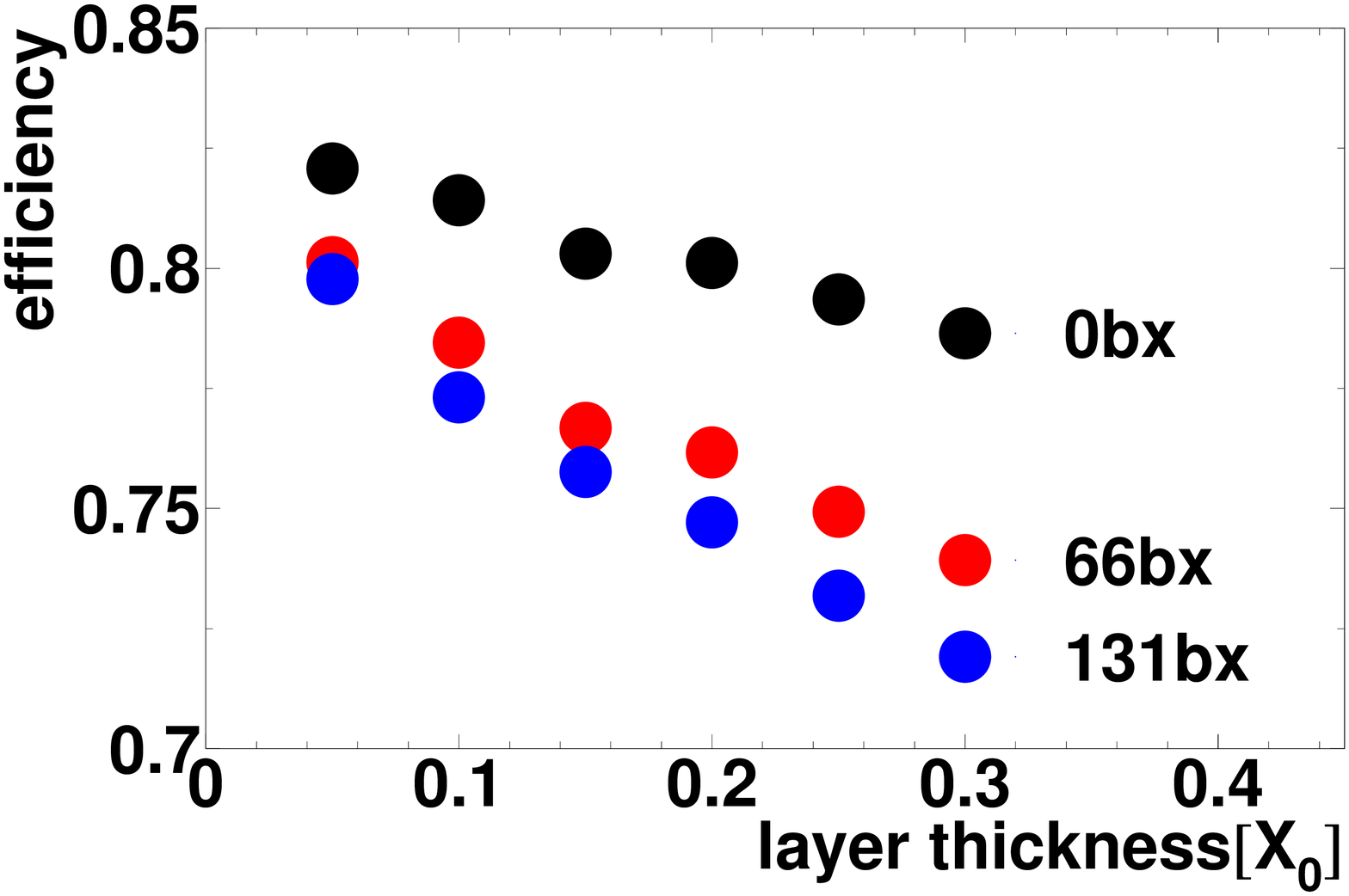}
\begin{minipage}{2mm} \vspace*{-5.5cm} e)\end{minipage}
\includegraphics[width=0.32\columnwidth]{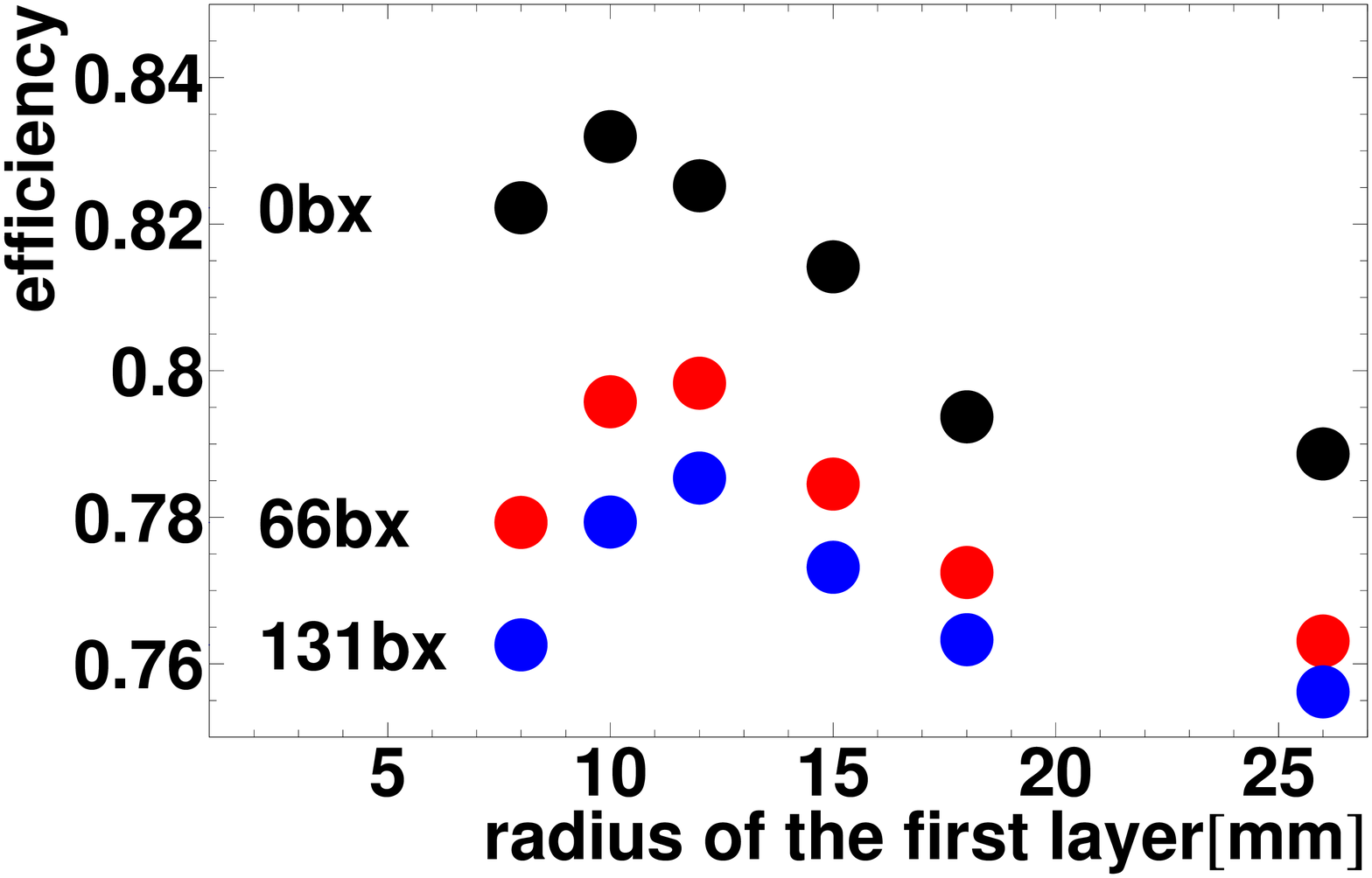}
\begin{minipage}{2mm} \vspace*{-5.5cm} f)\end{minipage}
\includegraphics[width=0.32\columnwidth]{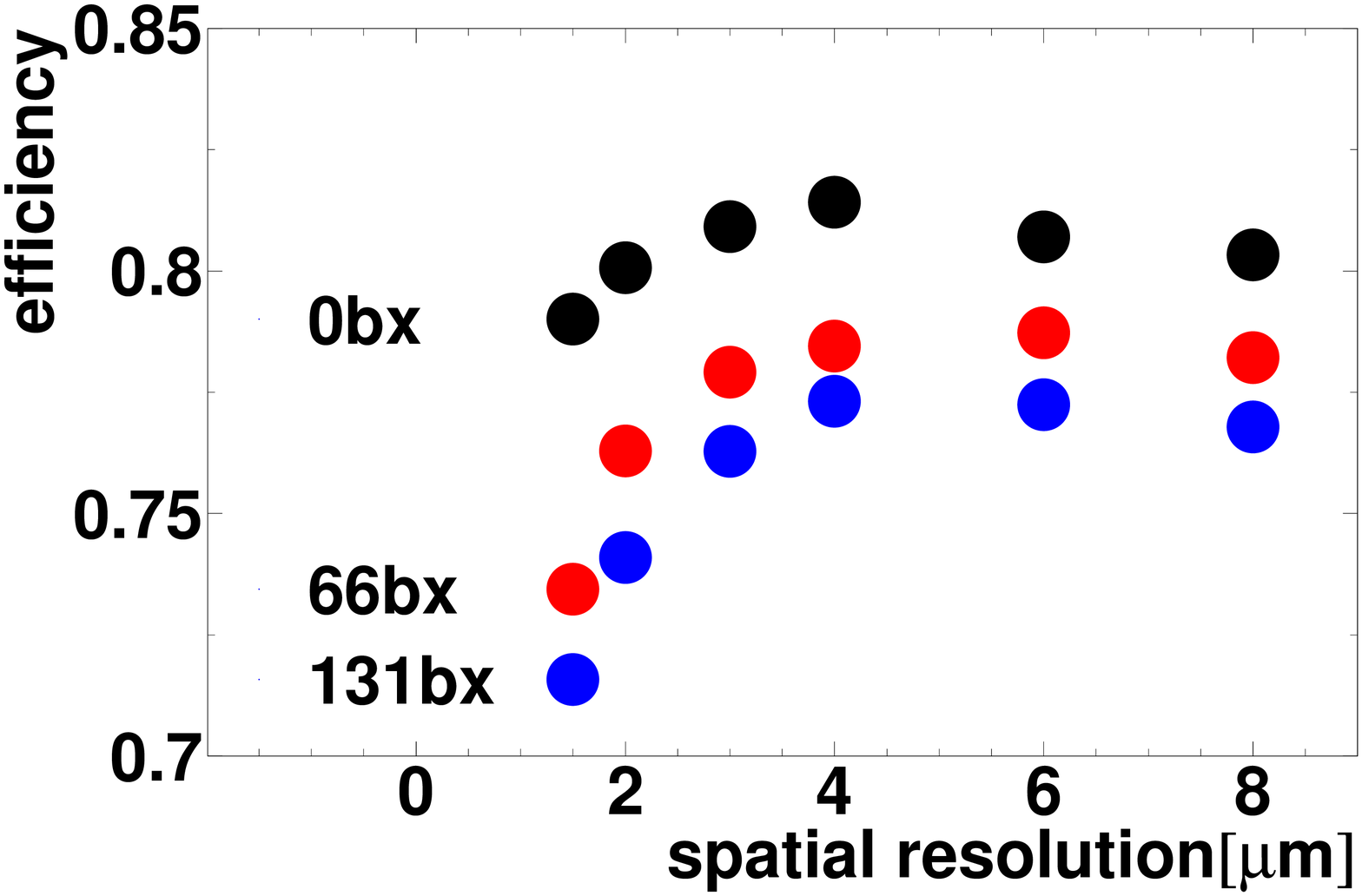}
}
\caption{$b$ jet selection: a), b), c) purity at fixed efficiency of 0.8;
 d), e), f) efficiency at~fixed~purity~of~0.8 for different sets of VTX geometries. }\label{pureffb}
\end{figure}

\begin{figure}[h]
\centerline{
\begin{minipage}{2mm} \vspace*{-5.5cm} a)\end{minipage}
\includegraphics[width=0.32\columnwidth]{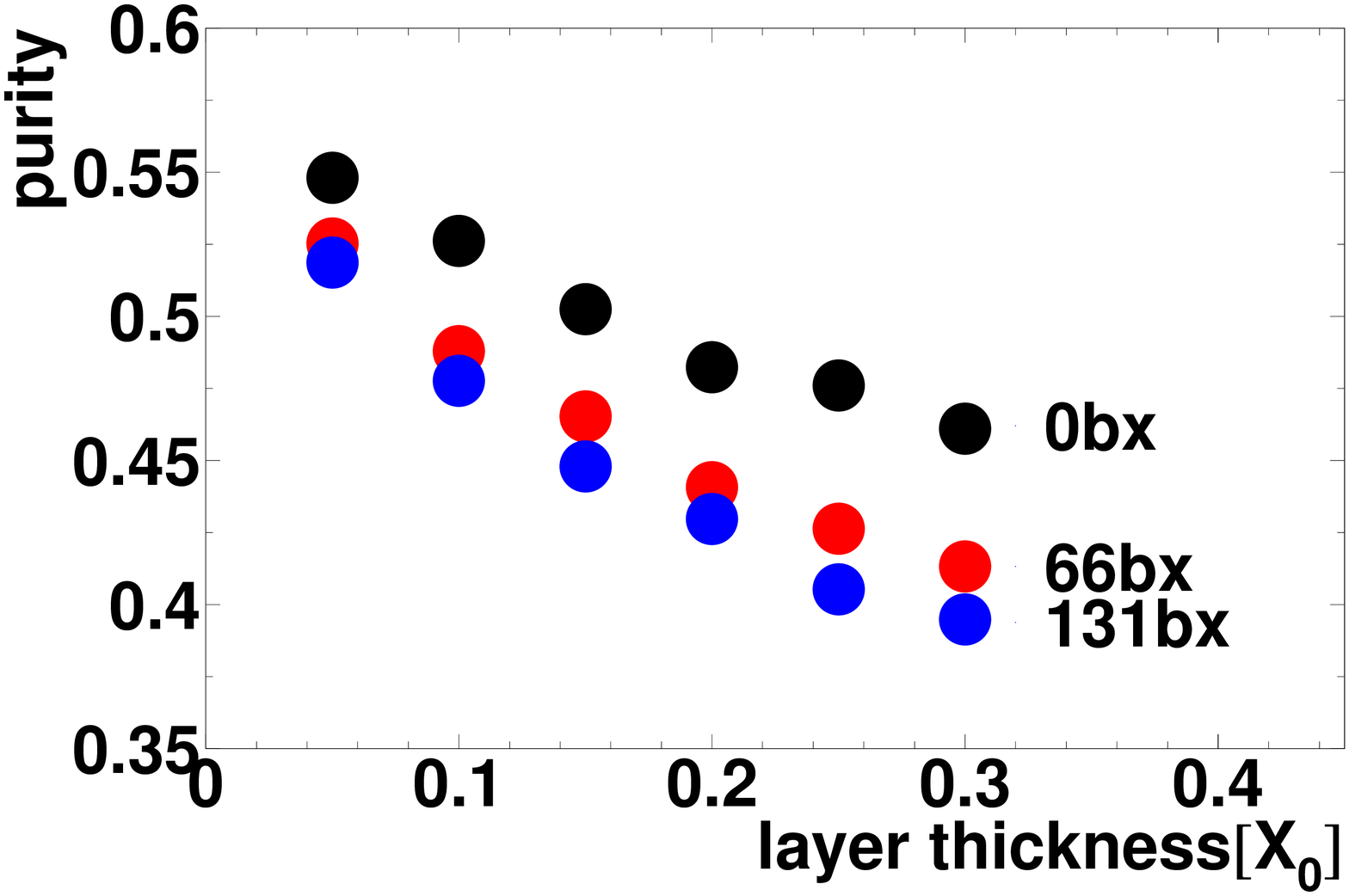}
\begin{minipage}{2mm} \vspace*{-5.5cm} b)\end{minipage}
\includegraphics[width=0.32\columnwidth]{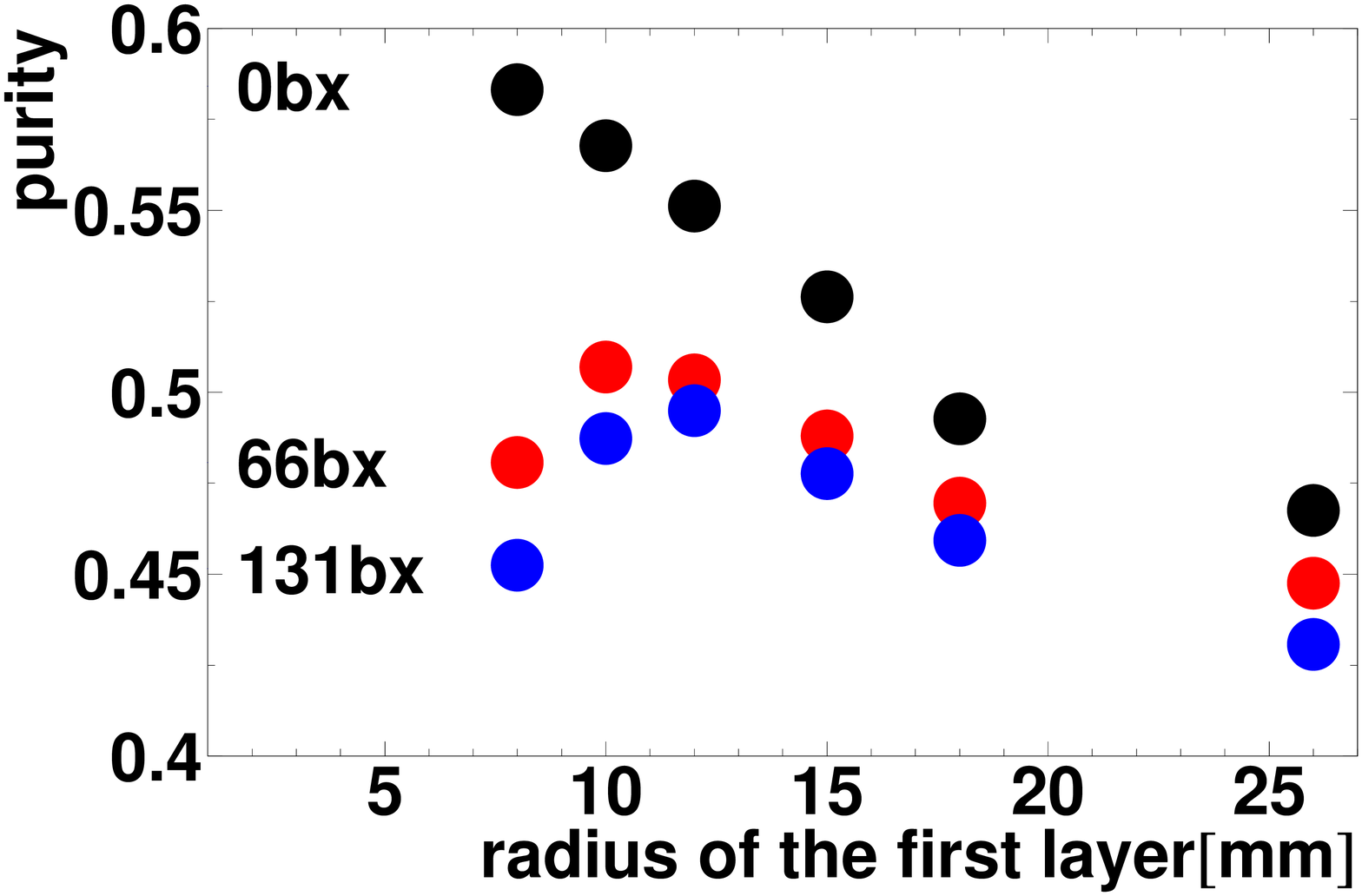}
\begin{minipage}{2mm} \vspace*{-5.5cm} c)\end{minipage}
\includegraphics[width=0.32\columnwidth]{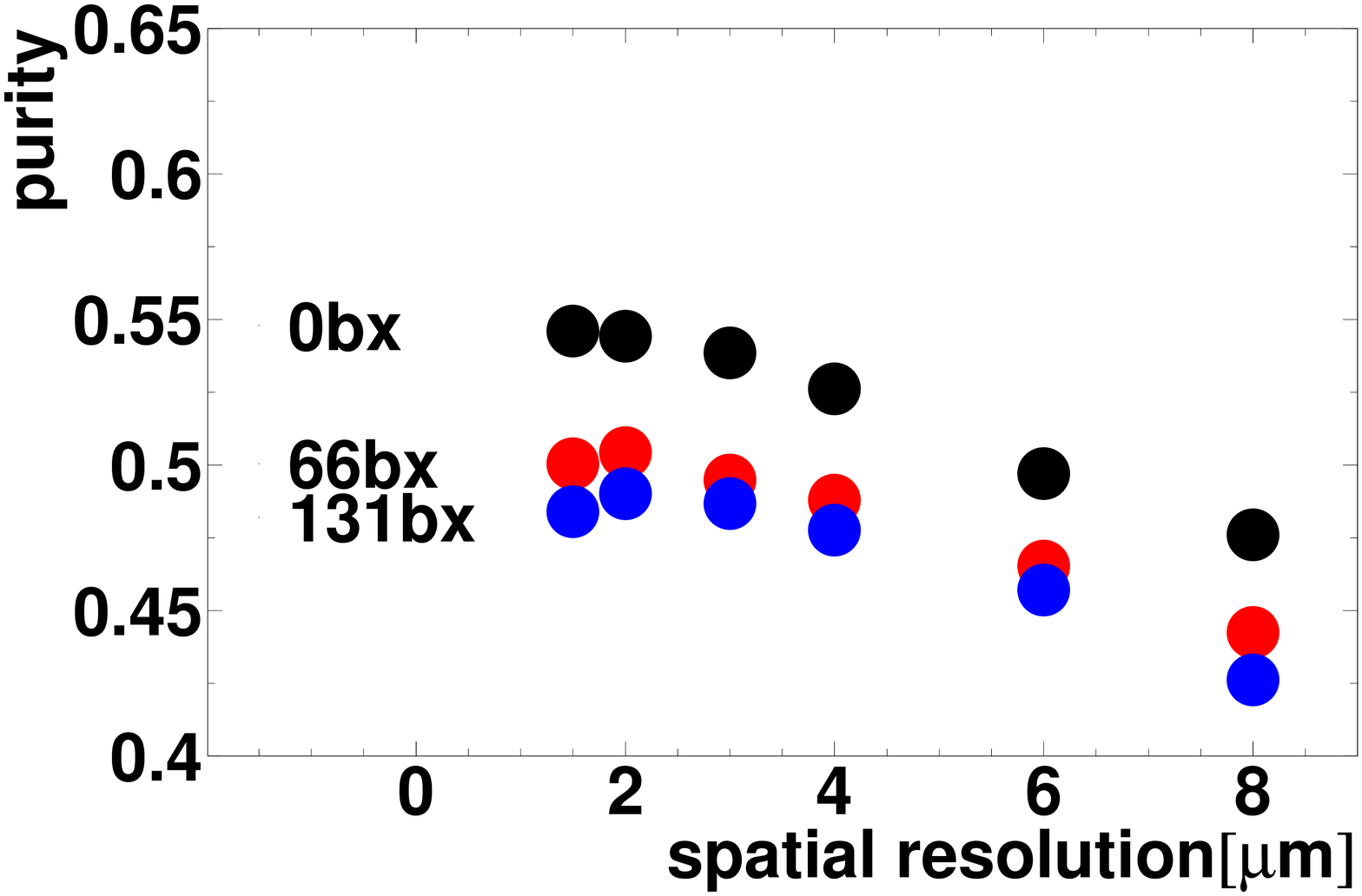}
}
\centerline{
\begin{minipage}{2mm} \vspace*{-5.5cm} d)\end{minipage}
\includegraphics[width=0.32\columnwidth]{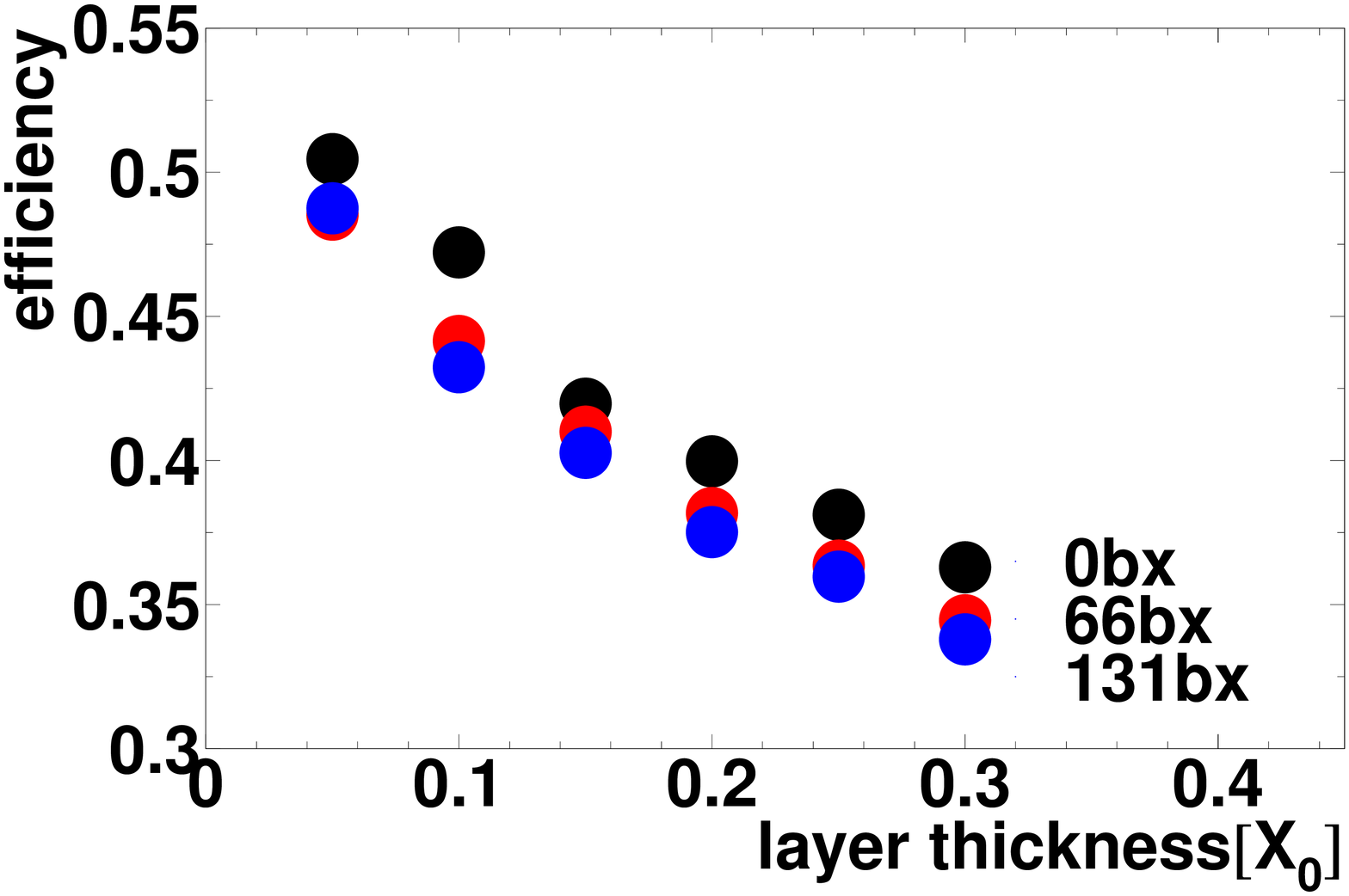}
\begin{minipage}{2mm} \vspace*{-5.5cm} e)\end{minipage}
\includegraphics[width=0.32\columnwidth]{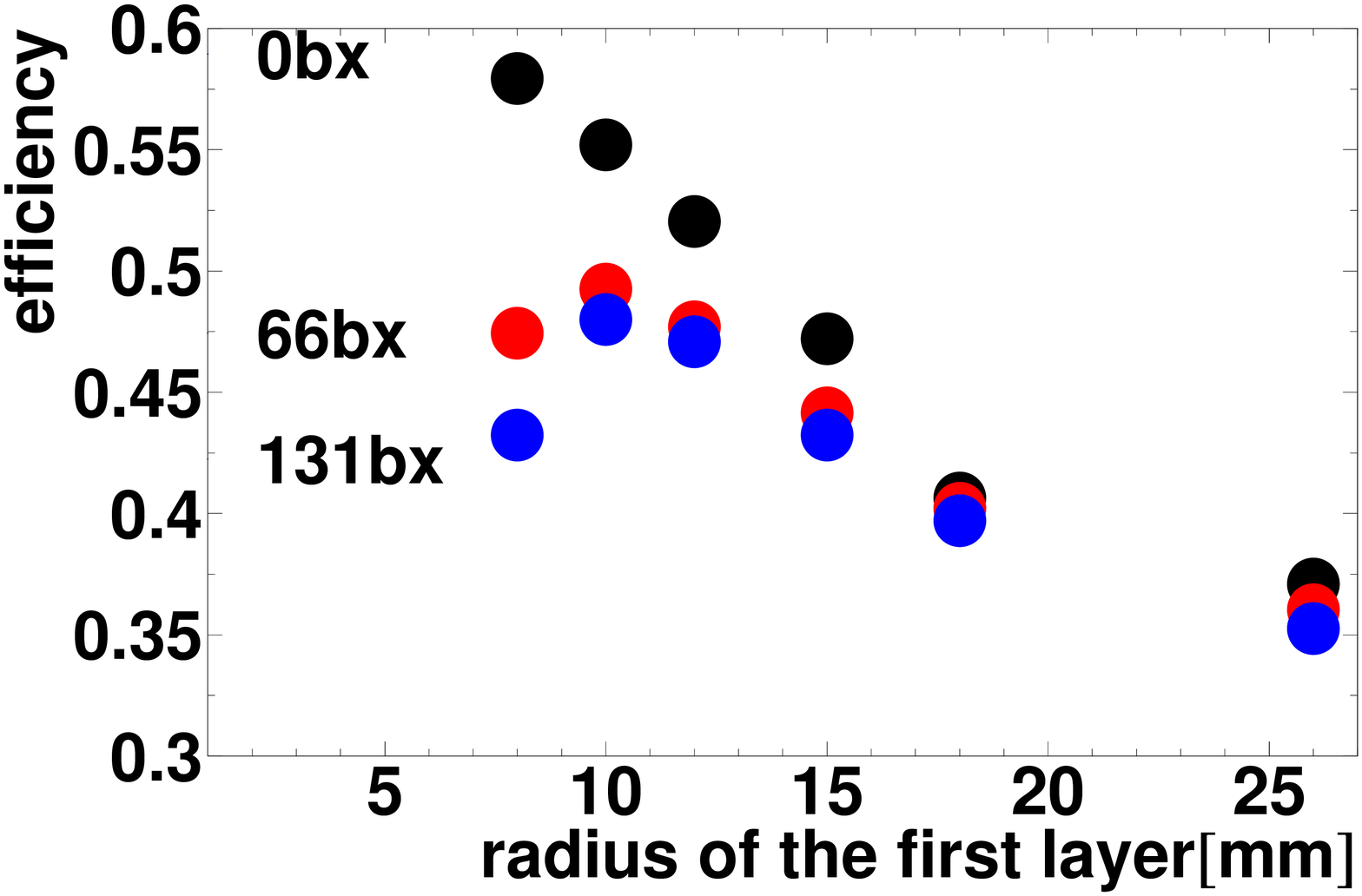}
\begin{minipage}{2mm} \vspace*{-5.5cm} f)\end{minipage}
\includegraphics[width=0.32\columnwidth]{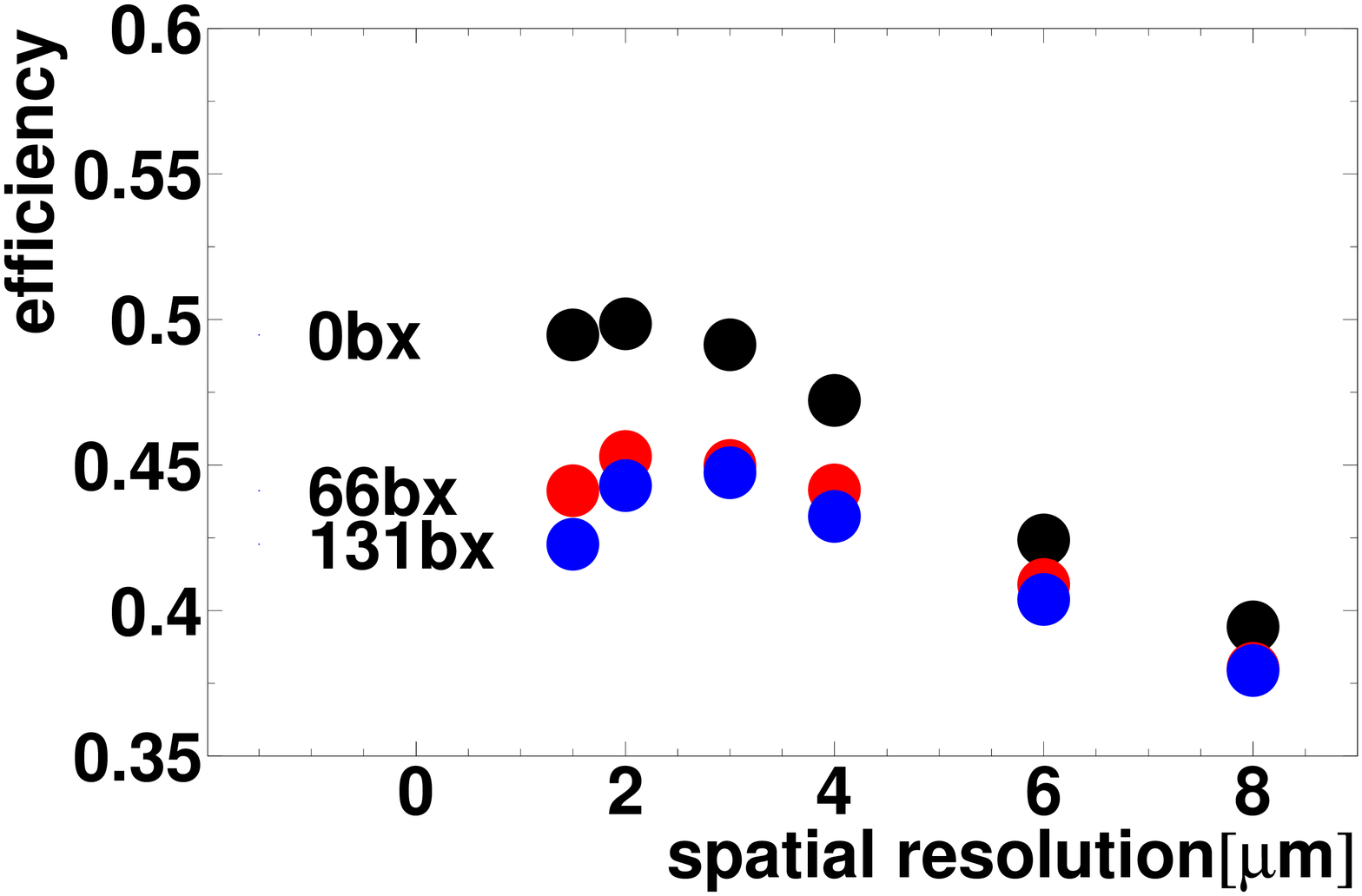}
}
\caption{$c$ jet selection: a), b), c) purity at fixed efficiency of 0.6;
 d), e), f) efficiency at~fixed~purity~of~0.6 for different sets of VTX geometries. }\label{pureffc}
\end{figure}

Jet flavour tagging performance was tested with 45.6 GeV jets in
the presence of the $e^+e^-$ background. Scenarios with 131BX,
66BX and no background were studied. For each scenario the
efficiency of selection at fixed purity and the purity of the
sample at fixed efficiency were studied. Results for $b$ and $c$
jet selection are shown on
Figures~\ref{pureffb}~and~\ref{pureffc}. Reduced layer thickness
improves jet flavour tagging performance and~reduces the impact of
the $e^+e^-$ background on jet flavour tagging. In the presence of
$e^+e^-$ background, the radius of the first layer should not be
smaller then 12 mm, otherwise jet flavour tagging performance
is~reduced. Spatial resolution better than 4 $\mu$m reduces
performance of jet flavour tagging due to increased probability
of~reconstructing fake secondary vertices, especially in presence
of $e^+e^-$ background.


\section{Measurement of the Higgs boson branching ratios}
\begin{figure}[h]
\centerline{
\begin{minipage}{2mm} \vspace*{-8.5cm} a)\end{minipage}
\includegraphics[width=0.3\columnwidth]{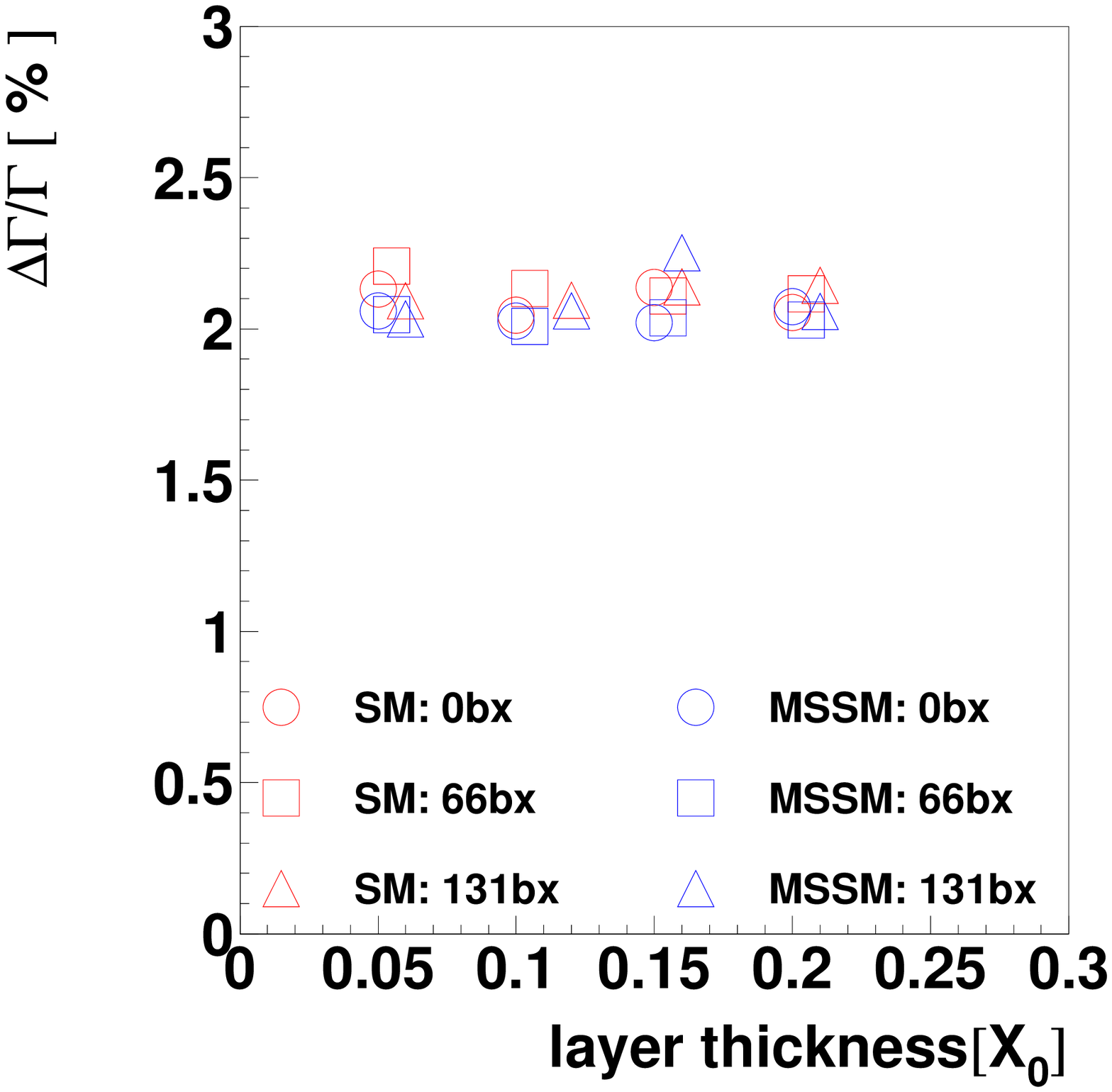}
\begin{minipage}{2mm} \vspace*{-8.5cm} b)\end{minipage}
\includegraphics[width=0.3\columnwidth]{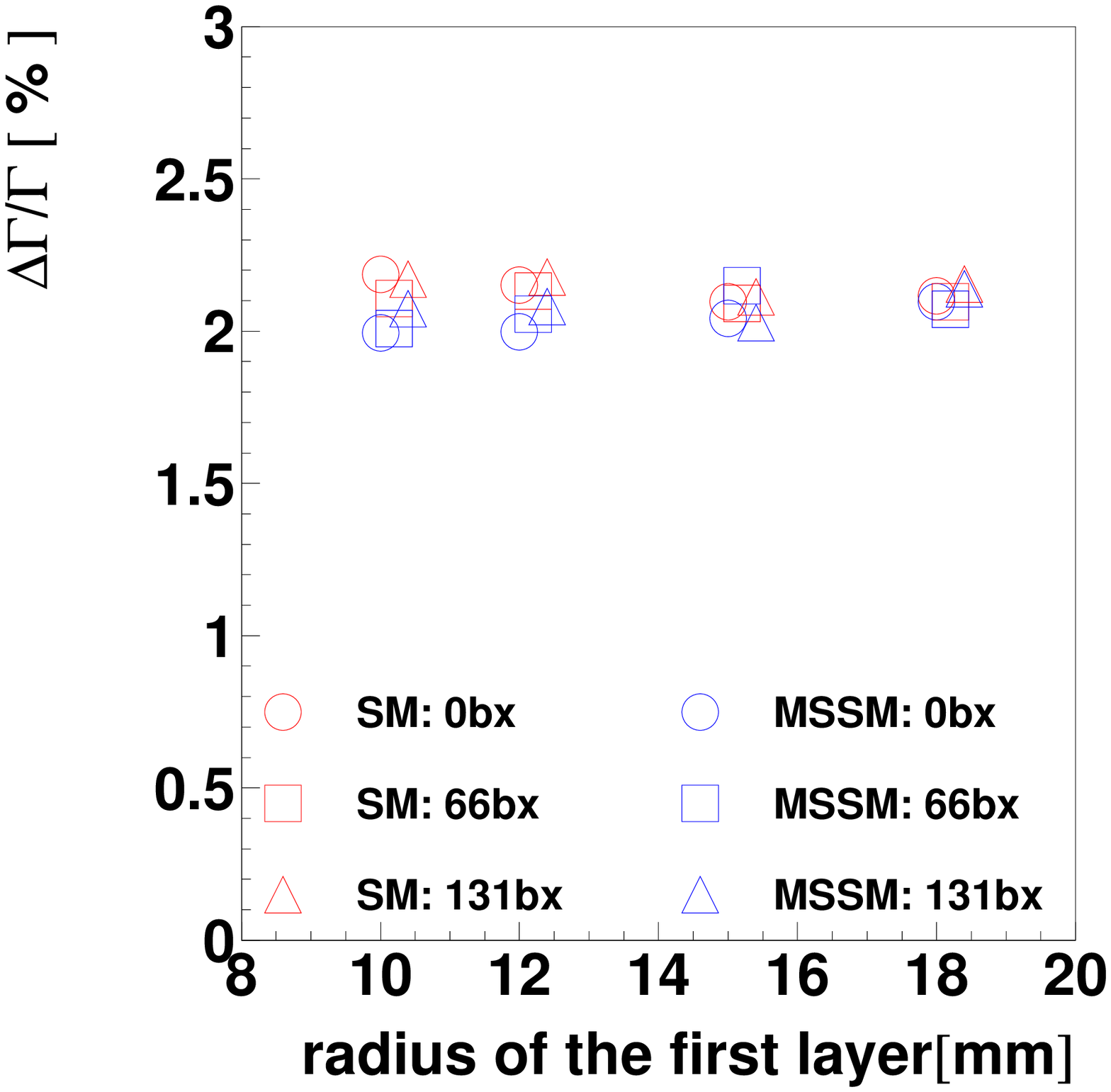}
\begin{minipage}{2mm} \vspace*{-8.5cm} c)\end{minipage}
\includegraphics[width=0.3\columnwidth]{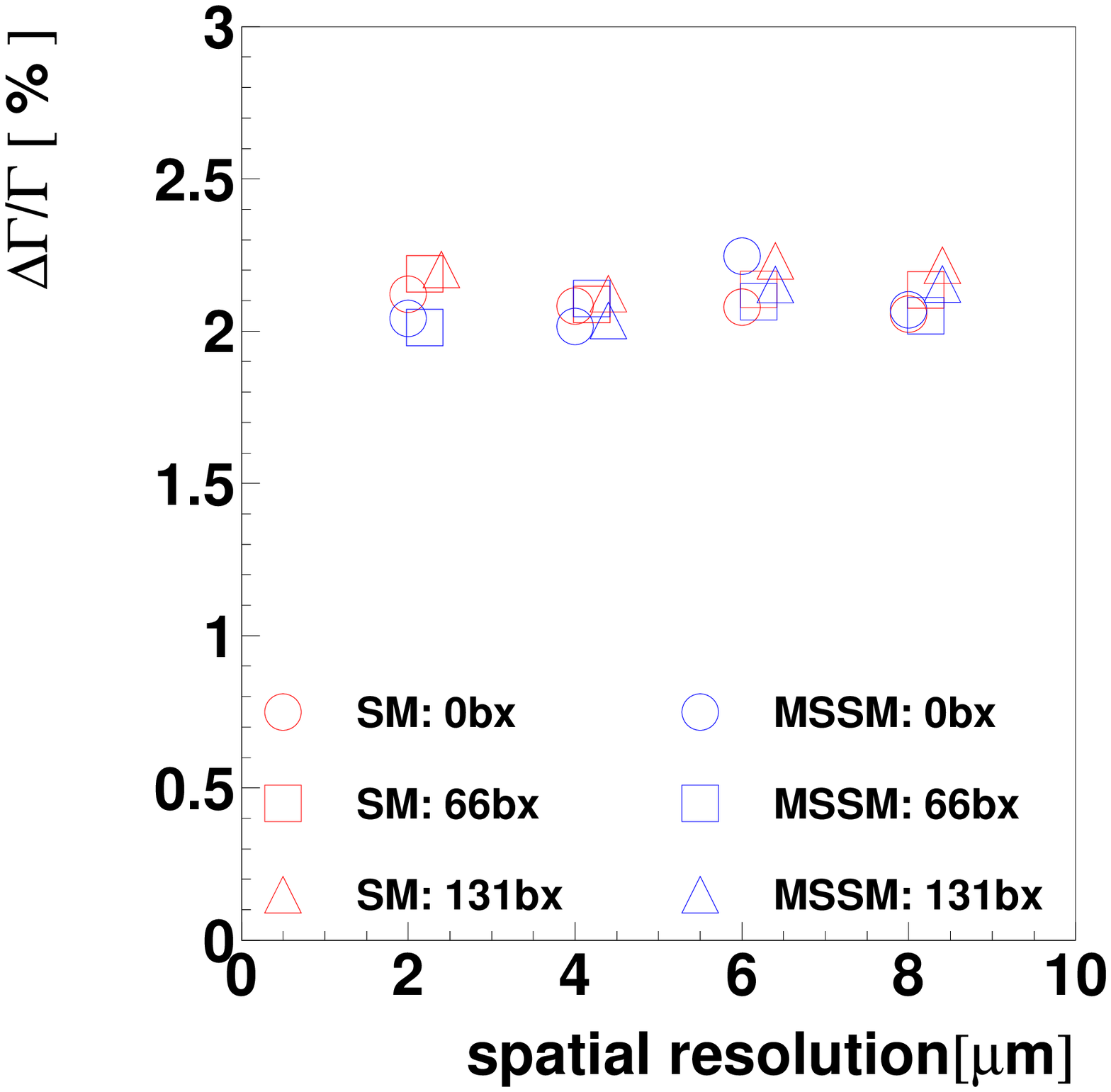}
\rotatebox{90}{\hspace{1.5cm} $H \rightarrow b\bar{b}$}
}
\centerline{
\begin{minipage}{2mm} \vspace*{-8.5cm} d)\end{minipage}
\includegraphics[width=0.3\columnwidth]{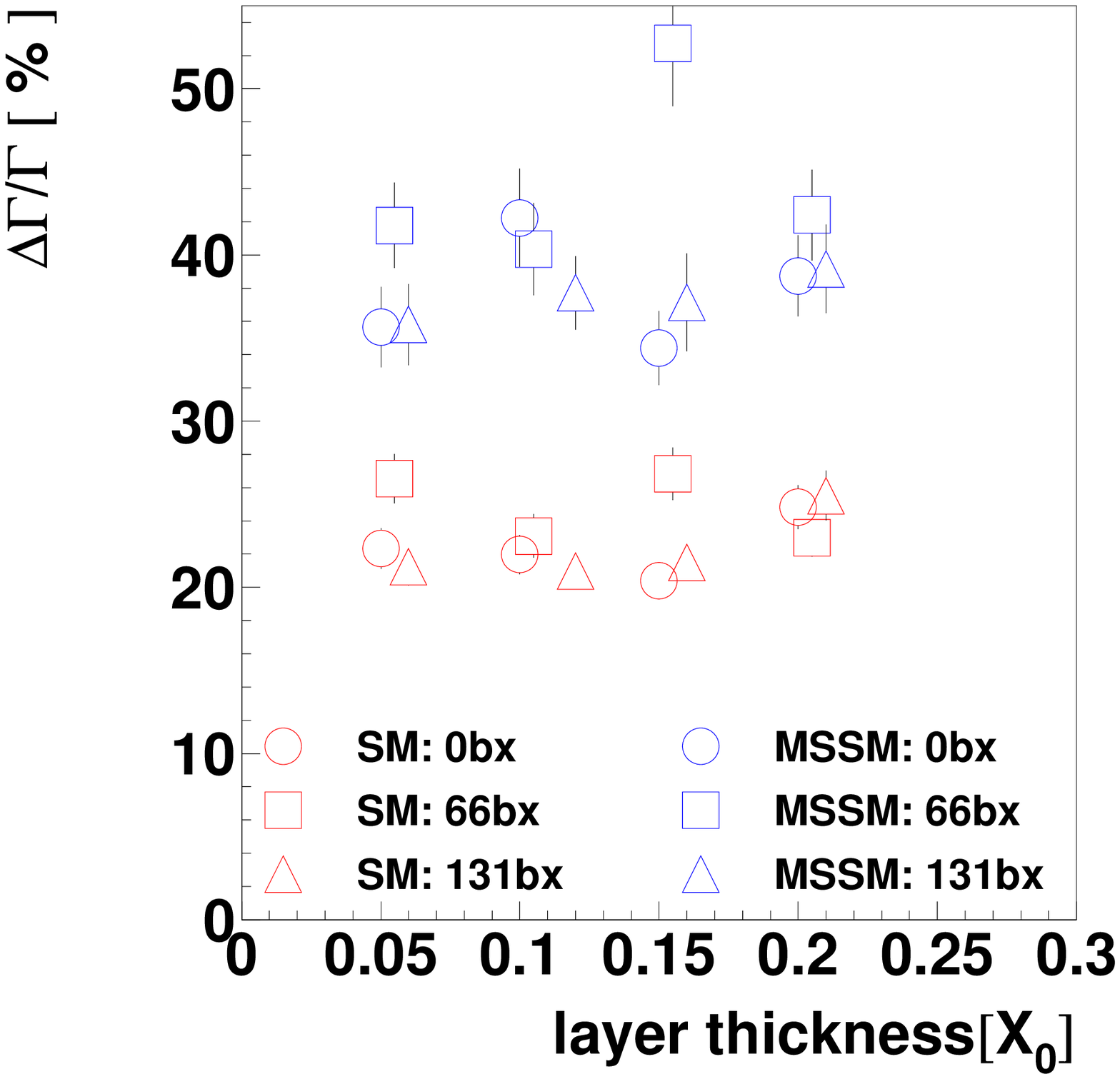}
\begin{minipage}{2mm} \vspace*{-8.5cm} e)\end{minipage}
\includegraphics[width=0.3\columnwidth]{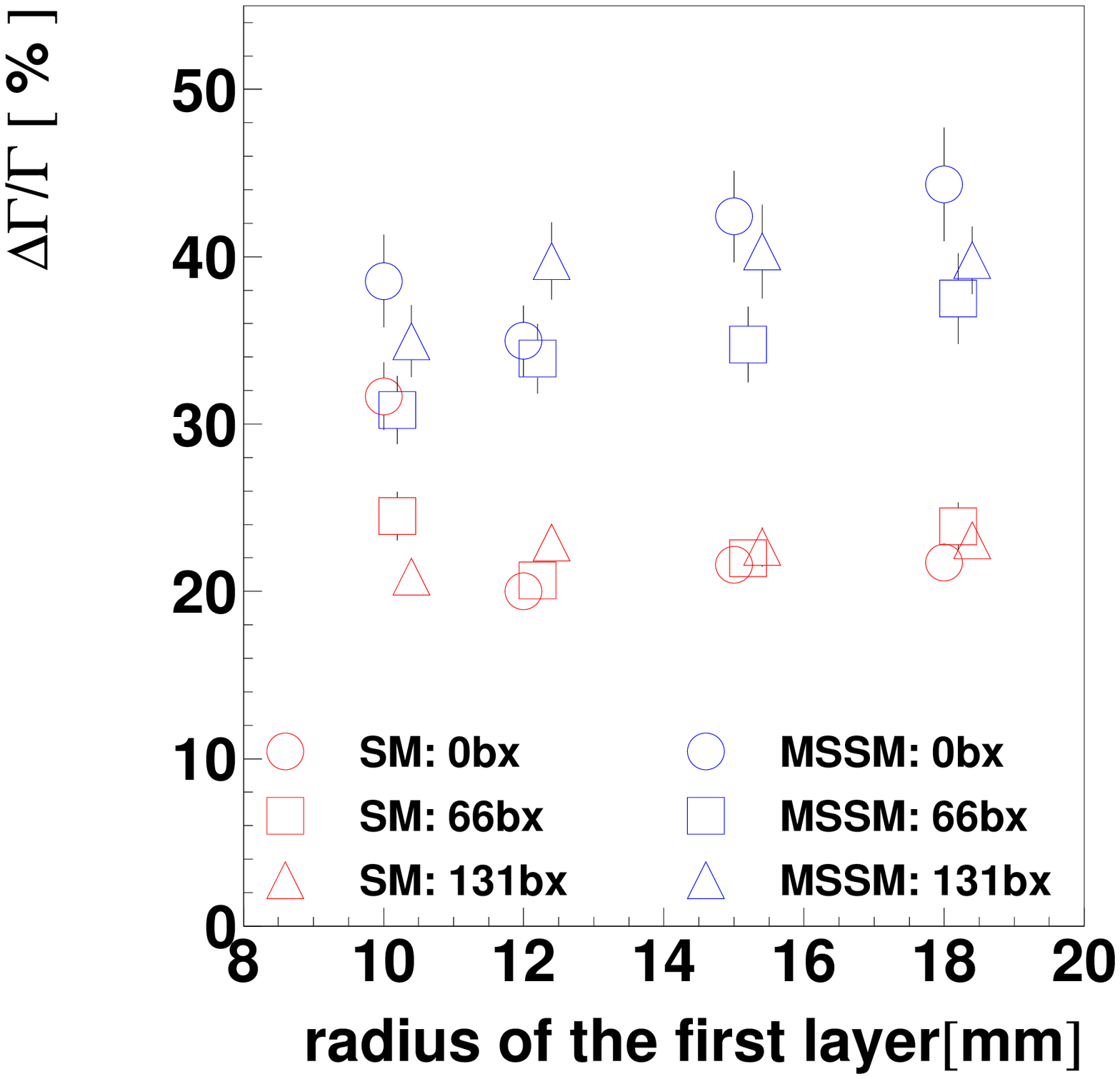}
\begin{minipage}{2mm} \vspace*{-8.5cm} f)\end{minipage}
\includegraphics[width=0.3\columnwidth]{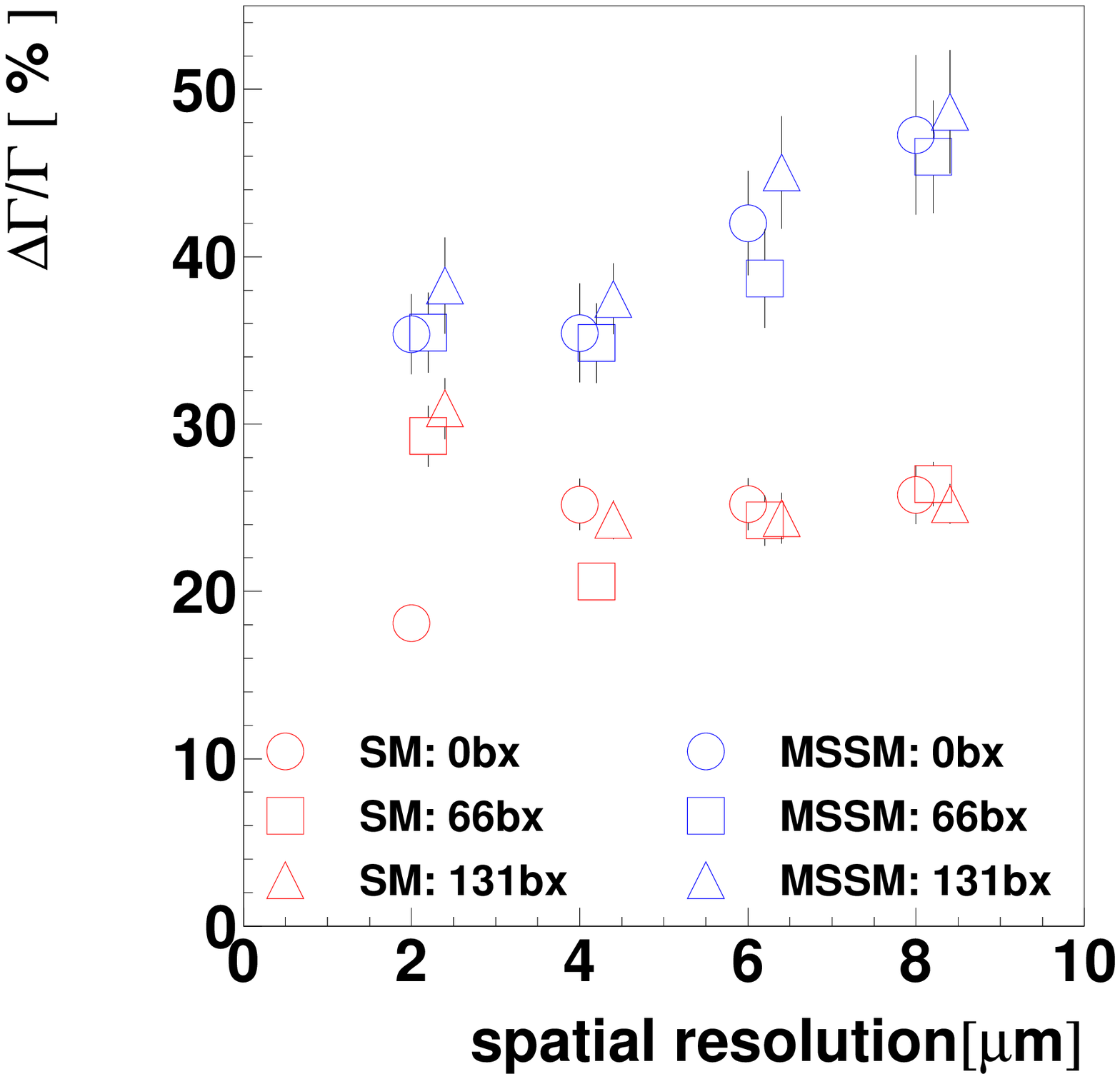}
\rotatebox{90}{\hspace{1.5cm} $H \rightarrow c\bar{c}$}
}
\vspace{-0cm}
\caption{Precision of measurement of Higgs boson branching ratio for different sets of the VTX parameters.}\label{Higgs_res}
\end{figure}

Measurement of the SM and MSSM Higgs boson branching ratios was
performed using simulated data for $M_h=127$~GeV,
$\sqrt{s}=500$~GeV and ${\cal L}=500$~fb$^{-1}$ in presence of
$e^+e^-$ background. The Higgs boson decay widths were calculated
using HDECAY \cite{hdecay}. The SM background processes $e^+ e^-
\rightarrow W^+ W^-$, $e^+ e^- \rightarrow q \bar{q}$, $e^+ e^-
\rightarrow Z Z$ were taken into account. Results are presented in
Figure~\ref{Higgs_res}. No significant influence of VTX parameters
on measurement of Higgs boson branching ratios is visible.
\begin{footnotesize}



%

\end{footnotesize}


\end{document}